\documentclass[aps,prl,twocolumn,amsmath,amssymb,showpacs,%
superscriptaddress,floatfix,]{revtex4}

\usepackage{graphicx}
\usepackage{epstopdf}

\newcommand{\Op}[1]{\boldsymbol{\mathsf{\hat{#1}}}}

\newcommand{\Fkt}[1]{\,\mathsf {#1}}

\def\openone{\leavevmode\hbox{\small1\kern-3.3pt\normalsize1}}

\begin{document}

\title{Pump-probe spectroscopy of two-body correlations in ultracold gases}
\date{\today}

\author{Christiane P. Koch}
\email{ckoch@physik.fu-berlin.de}
\affiliation{Institut f\"ur Theoretische Physik,
  Freie Universit\"at Berlin,
  Arnimallee 14, 14195 Berlin, Germany}
\author{Ronnie Kosloff}
\affiliation{Institute of Chemistry and
  The Fritz Haber Research Center, 
  The Hebrew University, Jerusalem 91904, Israel}

\date{\today}

\begin{abstract}
We suggest pump-probe spectroscopy to study pair
correlations that determine the many-body dynamics in weakly
interacting, dilute ultracold gases. 
A suitably chosen, short laser pulse depletes the pair
density locally, creating a 'hole' in the electronic
ground state. The dynamics of this non-stationary pair density
is monitored by a time-delayed probe pulse. The resulting 
transient signal allows to spectrally decompose the 'hole' and
to map out the pair correlation function. 
\end{abstract}

\maketitle

\paragraph*{Introduction}

Bose-Einstein condensation in dilute gases is determined by the nature
of the two-body interactions between the atoms \cite{PethickSmith}. These
microscopic interactions manifest themselves in two-body correlations
and dictate the mesoscopic and macroscopic properties of the
condensate.
Formally, the dynamics of an ultracold gas is described in terms of
field equations \cite{PethickSmith}. 
For dilute gases where only two-body interactions are prominent,
the equation of motion for the field operator that 
annhilates (or creates) a particle at position $\vec{x}$ reads
\begin{eqnarray}
  \label{eq:Heisenberg}
  i\hbar\frac{\partial \Op{\psi}}{\partial t}(\vec{x};t) &=&
  \Op{H}_1\Op{\psi}(\vec{x};t) + \\ \nonumber &&\int d^3\vec{y}
  \Op{V}_{\boldsymbol{2}}(\vec{x}-\vec{y}) \Op{\psi}^\dagger(\vec{y};t)
  \Op{\psi}(\vec{y};t) \Op{\psi}(\vec{x};t)\,.
\end{eqnarray}
Expectation values of the many-body system can be expressed in terms of
normal-ordered correlation functions. To lowest order, these 
are the condensate or mean field wavefunction,
$\Psi(\vec{x};t)=\langle \Op{\psi}(\vec{x};t)\rangle$, the one-body
density matrix $R(\vec{x},\vec{y};t)=\langle
\Op{\psi}^\dagger(\vec{x};t)\Op{\psi}(\vec{y};t)\rangle$ and the pair
correlation function,
$\Phi(\vec{x},\vec{y};t)=\langle\Op{\psi}(\vec{x};t)\Op{\psi}(\vec{y};t)\rangle$.  
For practical calculations, 
the infinite set of equations of motion for the many-body problem needs
to be truncated. This can achieved
by expanding the correlation functions into cumulants
\cite{KoehlerPRA02}. 
A separation of time or length
scales, i.e. small collision time vs long free propagation time or
small effective range of the interaction potential vs large
interatomic distance is required to justify truncation. 
For dilute Bose gases in a macroscopic trap,
such an assumption can typically be made, and within the first-order
cumulant expansion, the dynamics of pair correlations is decoupled
from higher order terms \cite{KoehlerPRA02}. 
Alternatively, one can work with the
correlation functions directly \cite{PascalPRA03}. 
In both cases, the dynamics of the macroscopic pair correlation
function is described by a Schr\"odinger-like equation where the
mean field enters as a source term \cite{KoehlerPRA02} or acts as an
additional potential \cite{PascalPRA03}. If we restrict our
considerations to timescales that are much shorter than the
timescale of the mean field dynamics, the pair correlation dynamics
are described by a standard Schr\"odinger equation where the 
presence of the condensate only modifies the boundary conditions. 
The macroscopic pair correlation
function is then given by the two-body wavefunction of an isolated pair
of atoms, $\Phi(r)$ with $r=|\vec{x}-\vec{y}|$. 
We can thus study the many-body pair correlation dynamics
by solving the time-dependent Schr\"odinger equation
for two colliding atoms \cite{PascalPRA03}.
\begin{figure}[b]
  \centering
  \includegraphics[width=0.95\linewidth]{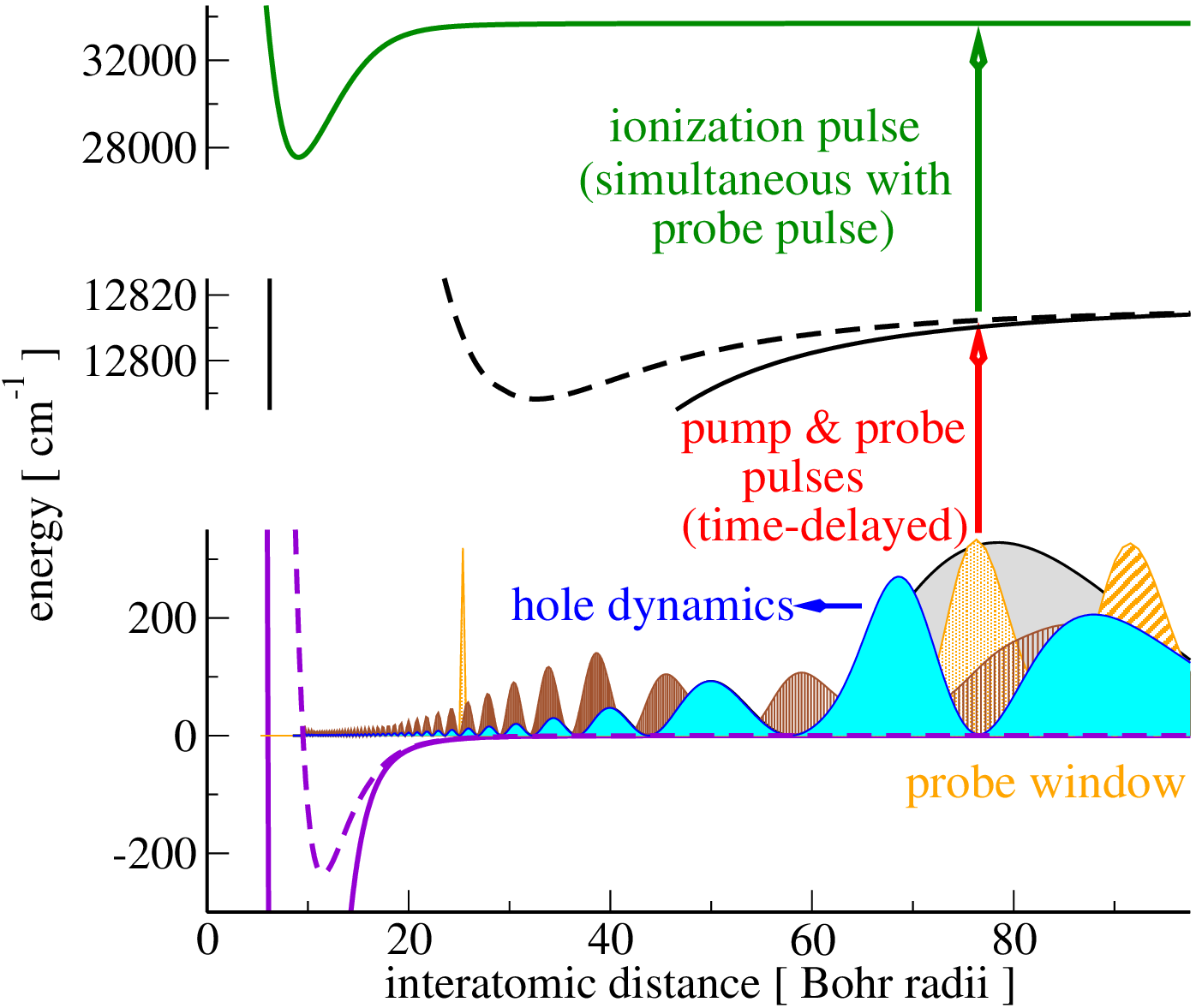}
  \caption{(Color online)
    Pump-probe spectroscopy of dynamical pair correlations:
    A pump pulse excites population from the electronic ground,
    leaving the pair correlation
    function in a non-stationary state, the 'hole'. 
    A time-delayed probe pulse monitors the dynamics of the 'hole'. 
    The orange peaks indicate the action where the probe pulse
    measures pair amplitude.
  }
  \label{fig:scheme}
\end{figure}

In the most simplified approach,
the effect of the many-body pair correlations
is captured in a single parameter, the scattering length
\cite{PethickSmith}. 
Measuring the scattering length corresponds to an indirect assessment
of the pair correlations. If the two-body interaction is probed in
a time much shorter than its characteristic timescale, 
a more comprehensive study becomes possible.
Here we suggest to employ  pump-probe
spectroscopy to unravel the dynamics of pair correlations 
in an ultracold Bose gas. This requires a 
combination of ultrafast and ultracold physics, the basic 
basic feasibility of which has been
demonstrated in recent experiments on femtosecond photoassociation of
ultracold rubidium atoms \cite{SalzmannPRL08,McCabePRA09}.

\paragraph*{Pump-probe spectroscopy}
Our scheme involves three short laser pulses and 
is sketched in Fig.~\ref{fig:scheme} for the example of an ultracold
gas of $^{87}$Rb atoms.
The pump pulse excites population from the electronic ground to an
excited state, leaving a 'hole' in the initial pair correlation
function. The 'hole' represents a
non-stationary state that moves under the influence of the
ground state potential, cf. Fig.~\ref{fig:scheme}.
The pump pulse thus induces the dynamics of the pair correlations.
A time-delayed probe monitors these dynamics by measuring
the amount of probability amplitude in a
range of internuclear distances. The measurement consists of applying
simultaneously a photoassociation and an ionization pulse (combination of red and
green arrows in Fig.~\ref{fig:scheme}). The pair density on the ground
state is thus photoassociated and immediately ionized for detection.  
The largest probe signal is obtained when the probe pulse is 
identical to the pump pulse. The dynamics are then 
monitored at the position where the 'hole' was created. The spatial
region where the probe pulse detects pair density is indicated 
in orange in Fig.~\ref{fig:scheme}.

For alkali atoms, the initial pair correlation function consists of a
superposition of singlet and triplet components. The corresponding
interaction potentials are shown in solid (singlet) and dashed
(triplet) lines. For clarity's sake, only triplet pair
wave functions are depicted in Fig.~\ref{fig:scheme}. Since the
respective excited state potentials differ, the probe pulse is
resonant at different interatomic distances 
for singlet and triplet. This is indicated by the orange
peaks representing the probe windows in Fig.~\ref{fig:scheme} (striped
for singlet, dotted for triplet).

Detection of the pair correlation dynamics 
proceeds via the creation of molecular ions and 
is inspired by Refs.~\cite{SalzmannPRL08,McCabePRA09}.
In those experiments, the pump pulse photoassociates ultracold atoms 
and the ensuing excited state molecular dynamics are detected by the
ionization pulse. In contrast to that, it is the ground state dynamics
of the many-body pair correlations that are probed in our proposal;
photoassociation just serves as a means for detection. Hence, while
the experimental setups envisioned in this work and realized in
Refs.~\cite{SalzmannPRL08,McCabePRA09} are similar, the probed physics 
is rather different. 

\paragraph*{Modelling the two-body interaction}
We consider two colliding $^{87}$Rb atoms. Hyperfine interaction
couples the ground state singlet and lowest triplet scattering
channels. However, this interaction cannot be resolved on the timescales
considered below. We therefore assume a superposition of singlet and
triplet components, but neglect the effect of hyperfine interaction on
binding energies and dynamics.
The two-body Hamiltonian is represented on a grid large
enough to faithfully represent the scattering atoms.
The interaction of the atom pair with the pump pulse is treated within 
the dipole and rotating wave approximations.
Excitation is considered
exemplarily into the $0_u^+(5s+5p_{3/2})$ and $0_g^-(5s+5p_{3/2})$
excited states.
The pulses are taken to be transform-limited Gaussian pulses
with a full-width at half-maximum (FWHM) of 10$\,$ps. This corresponds
to a spectral bandwidth of roughly $1.5$cm$^{-1}$ or 45$\,$GHz.
Details on the potentials and the employed methods are
found in Ref.~\cite{MyPRA06b}. 
For a Bose-Einstein condensate, a single low energy
scattering state needs to be considered. 
The collision energy of this initial state 
is chosen to correspond to 20$\,\mu$K with 75\% (25\%) triplet
(singlet) character. 
At higher temperatures, the bosonic nature of the atoms can be
neglected; and the ultracold thermal ensemble is described by a
Boltzmann average over all thermally populated two-body scattering
states \cite{MyJPhysB06}.

\paragraph*{Modelling the absorption of the probe pulse}
The dynamics of the non-stationary 'hole' 
is monitored by a combination of a probe pulse
and  an ionization pulse, cf. Fig.~\ref{fig:scheme}.
This two-color scheme converts the absorption of the probe pulse into
detection of molecular ions.
Assuming the probe pulse to be weak and the ionization step
to be saturated, absorption of the probe pulse can be modelled within
first order perturbation theory~\cite{cina97,Jiri00}.
The transient absorption signal is then represented by the
time-dependent expectation value of a window operator,
\begin{equation}
  \label{eq:window}
\Op{W}(\Op{r}) =
\pi (\tau_p E_{p,0})^2 \Fkt{e}^{-2 \Op{\Delta}(\Op{r})^2\tau_p ^2}\cdot 
\Op{\mu}^2  \,,
\end{equation}
where $\tau_p$ and
$E_{p,0}$ denote duration (FWHM) and peak amplitude of the probe pulse.
$\Op{\mu}_p$ is the transition dipole moment between ground and first
excited state. The central frequency of the probe pulse, $\omega_p$,
determines the difference potential, 
$\Op{\Delta}(\Op{r}) = V_{e}(\Op{r}) - V_\mathrm{g}(\Op{r}) - \hbar \omega_p$. 
A position measurement
becomes possible if the difference between the ground and
excited state potential is sufficiently large, and, moreover, the
spectral bandwidth sufficiently small to probe only non-zero
$\Op{\Delta}(\Op{r})$. Since the difference potential vanishes for
$r\to \infty$, this implies sufficiently detuned, narrow-band probe
pulses. Fig.~\ref{fig:scheme} shows
triplet and singlet window operators (orange peaks)
assuming identical parameters for pump and probe pulses.

\paragraph*{Characterization of the 'hole'}
As sketched in Fig.~\ref{fig:scheme}, the pump pulse carves a 'hole'
into the ground state pair correlation function. The resulting
non-stationary wave packet is a superposition of a few weakly bound
vibrational wave functions and many scattering states \cite{ElianePRA07}.
The detuning of the pump pulse from the atomic resonance frequency,
$\Delta_L=\omega_L-\omega_{at}$,
determines the position where the 'hole' is created:
For larger detuning, excitation occurs at shorter distance and
populates deeper bound levels.
A pulse energy of $\mathcal{E}_P=1.5\,$nJ is
sufficient to deplete the population within the resonance window of
the pump pulse (cf. the blue wave function in Fig.~\ref{fig:scheme}). At 
higher pulse energies, Rabi cycling within the resonance window sets
in. This leads to more population in the bound levels but also to
stronger redistribution among the scattering states.

\paragraph*{Dynamics of the 'hole'}
Fig.~\ref{fig:scheme} also illustrates the time evolution of the pair
correlation amplitude after a weak pump pulse
($\mathcal{E}_P=1.5\,$nJ) has been applied: The 
blue curve depicts the wave function just after the pump pulse,
at $t=24\,$ps (taking $t=0$ to be the time of the pump pulse 
maximum). A probe measurement at that time will find no amplitude
within the probe window. Due to the attractive interaction potential, 
the 'hole' moves toward shorter distances, cf. the brown
wave function ($t=126\,$ps). This brings
amplitude that is initially at larger 
$r$ into the probe window. Eventually the motion of the 'hole' will be
reflected at the repulsive barrier of the potential. The bound part of
the wave packet will remain at short distance and oscillate,
while the scattering part will pass through the probe window once not to
return.

\paragraph*{Pump-probe spectra}
The dynamics of the 'hole' is reflected in the transient probe
absorption, i.e. the time-dependent expectation value of the window
operator (red dotted curve in Fig.~\ref{fig:window} a):
\begin{figure}[tb]
  \centering
  \includegraphics[width=0.95\linewidth]{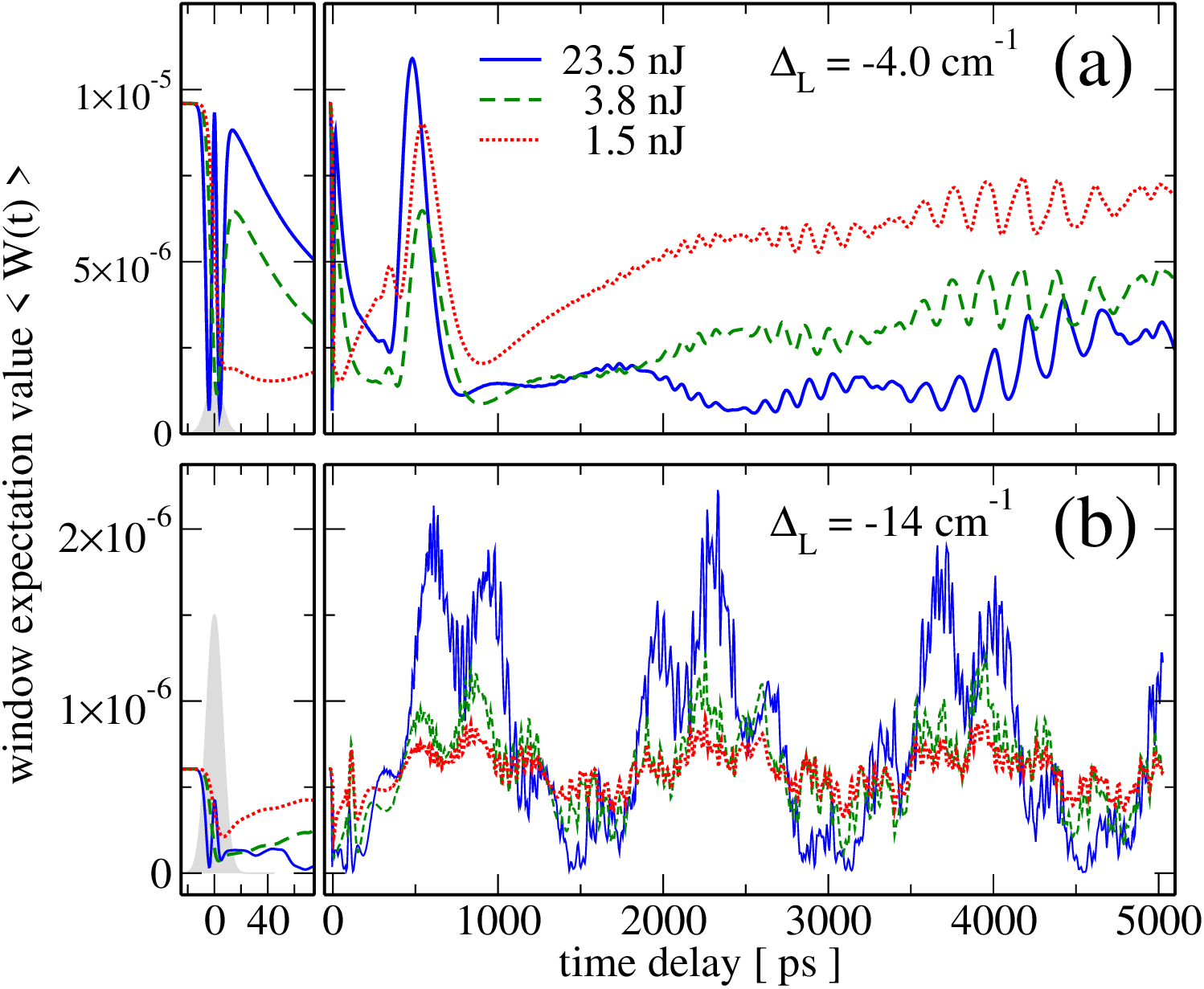}
  \caption{(Color online)
    Probing the two-body correlation dynamics: Absorption of the probe
    pulse as a function of the delay between pump and probe pulses. 
    The pump pulse detunings are $\Delta_\mathrm{L}=-4.0\,$cm$^{-1}$ (a) and 
    $\Delta_\mathrm{L}=-14\,$cm$^{-1}$ (b), and three different
    pump pulse energies are shown.
  }
  \label{fig:window}
\end{figure}
A depletion of the signal due to the creation of 
the 'hole' by the pump pulse, referred to as 'bleach' in traditional
pump-probe spectroscopy, is followed by a recovery that peaks at
550$\,$ps. At later times 
oscillations due to partial recurrence are observed but a full recovery
does not occur. Rabi cycling induced by larger pulse energies may
partially (green solid curve) or completely  (blue dashed
curve) refill the 'hole'.
For larger detuning, cf. Fig.~\ref{fig:window}b,
the 'hole' is created at shorter
interatomic distance, $r\sim 48\,$a$_0$ for $\Delta_L=-14\,$cm$^{-1}$
vs $r\sim 76\,$a$_0$ for $\Delta_L=-4\,$cm$^{-1}$. 
Obviously, the time to move to the repulsive barrier and back is then
shorter. A faster recovery of the
bleach is hence observed -- at $t=110\,$ps in Fig.~\ref{fig:window}b.

\begin{figure}[tb]
  \centering
  \includegraphics[width=0.95\linewidth]{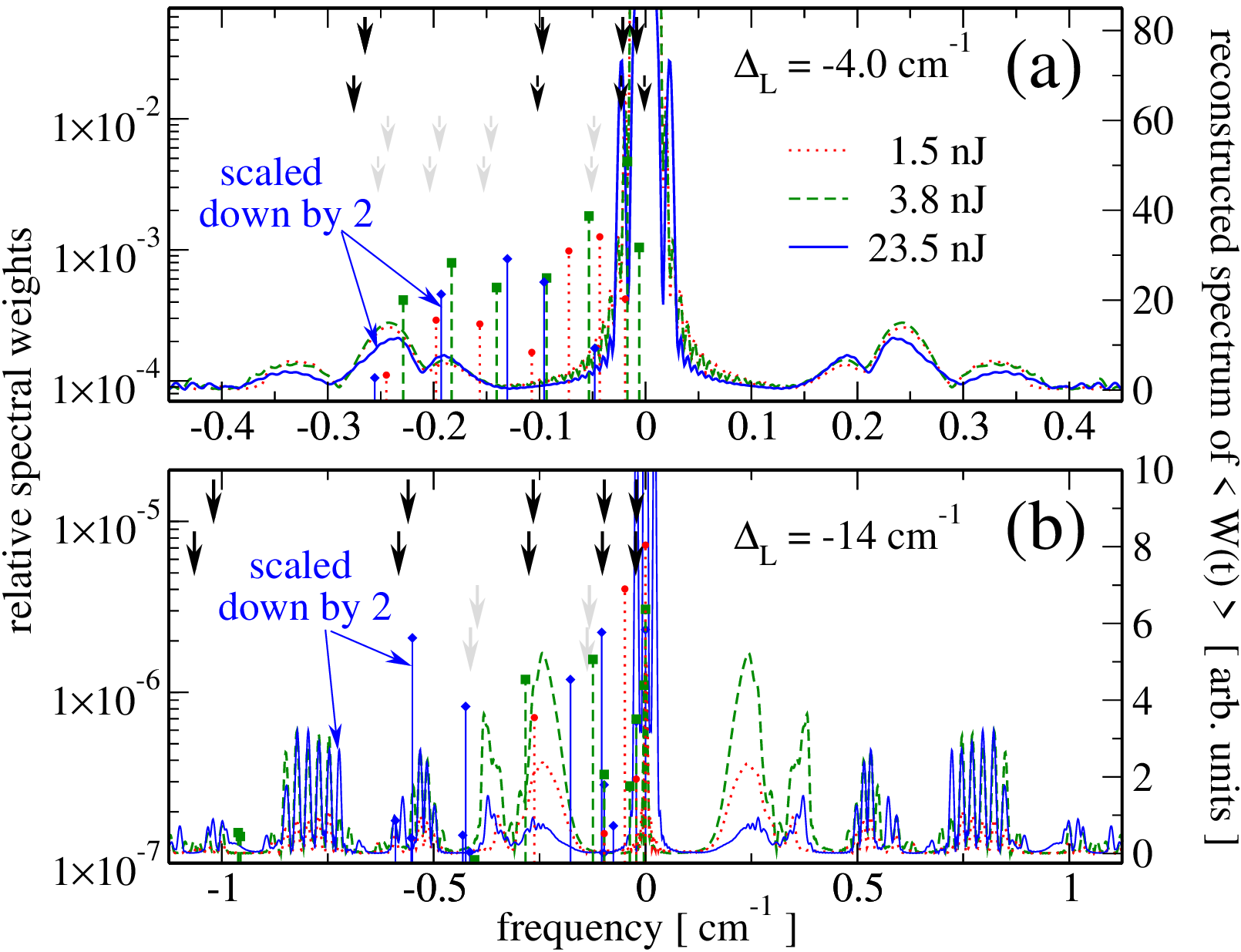}
  \caption{(Color online)
    Spectra of the transient absorption signals shown in
    Fig.~\ref{fig:window}. The eigenergies of the interaction
    potential are recovered:
    Black arrows indicate the position of the
    binding energies of the triplet (upper row) and singlet (lower row)
    levels, grey arrows half-multiples of the binding energies.
    The spectra for $23.5\,$nJ (blue solid lines) are scaled down by a
    factor of two. 
  }
  \label{fig:specwindow}
\end{figure}

The spectrum of the transient absorption signals shown in
Fig.~\ref{fig:window} can be 
obtained by filter-diagonalization \cite{MandelTaylor97}, 
a method allowing to accurately extract frequencies from just a few
oscillation periods. The spectra are shown in Fig.~\ref{fig:specwindow}.
The vibrational energies of the two-body interaction potential are
recovered. Fig.~\ref{fig:specwindow} shows furthermore
that Rabi cycling during the pump pulse leads to a larger bound part
in the 'hole', cf. the increase of spectral weights with pump pulse
energy. While a direct measurement of the vibrational populations 
would be difficult to implement experimentally, wave packet spectral
analysis via probe absorption is fairly straightforward.

\begin{figure}[tb]
  \centering
  \includegraphics[width=0.95\linewidth]{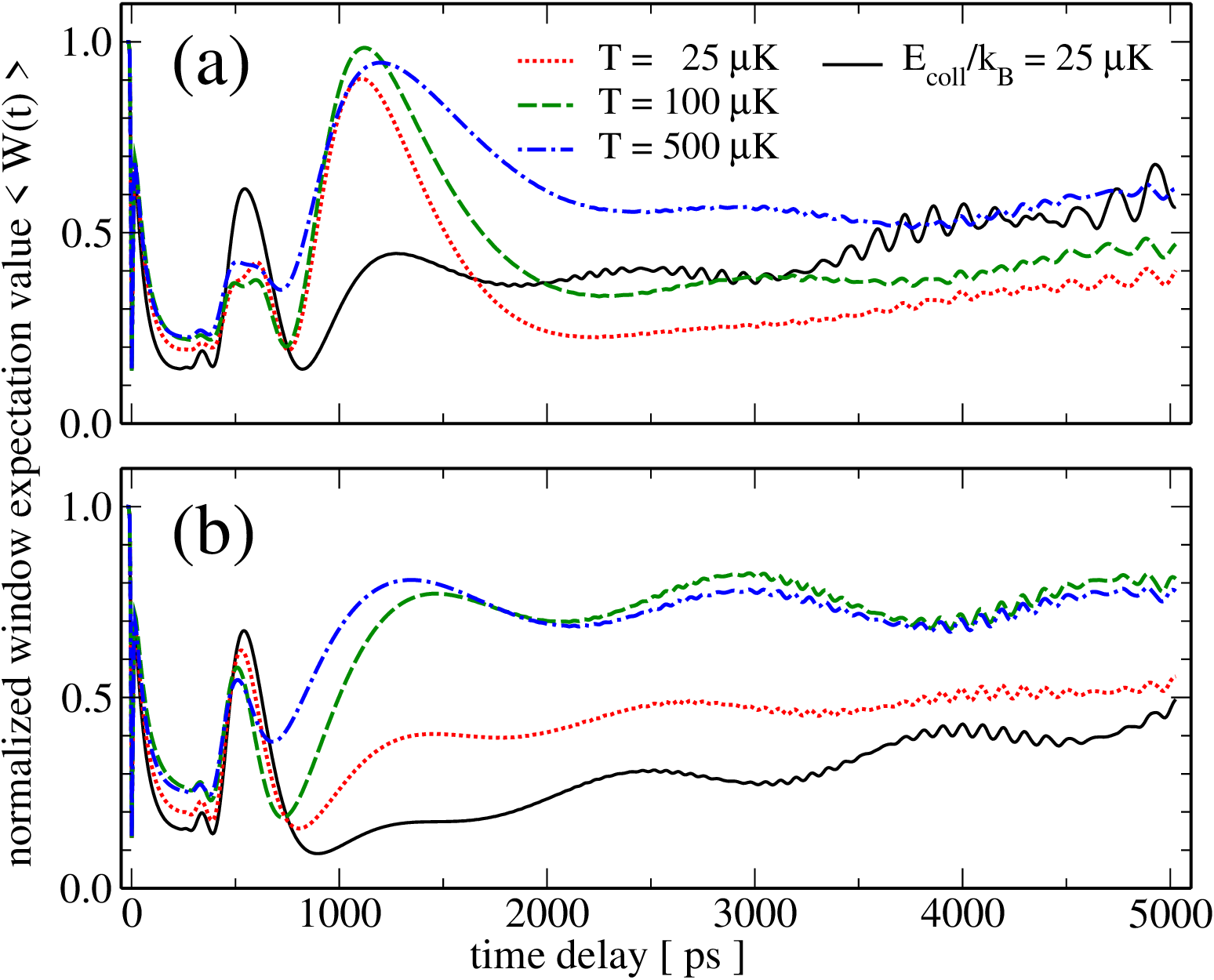}
  \caption{(Color online)
    Absorption of the probe pulse as a function of the delay between
    pump and probe pulses comparing 
    pure state dynamics (black solid lines) to those of a thermal
    ensemble (colored broken lines).
    (a) Calculation including two scattering channels (singlet and
    triplet): 
    The thermal dynamics are dominated by a singlet shape
    resonance while the pure dynamics shows features of both singlet
    and triplet dynamics. (b) Calculation for a single channel (triplet):
    The thermal and pure state dynamics are similar with the
    recovery of the bleach smeared out at higher temperatures.
    ($\Delta_\mathrm{L}=-4.0\,$cm$^{-1}$, $\mathcal{E}_P= 3.8\,$nJ)
  }
  \label{fig:thermav}
\end{figure}

\paragraph*{Pure state vs thermally averaged dynamics}
Pump probe spectroscopy of the pair correlations can be applied to
a condensate  as well as a thermal ultracold gas. For the
timescales considered here, this translates into comparing the
dynamics of a pure state to that of an incoherent
ensemble. Fig.~\ref{fig:thermav} shows the transient probe absorption
for the two cases. 
Thermal averaging markedly smears out the  recovery of
the bleach at 550$\,$ps. In principle,
thermal averaging introduces two effects -- the finite width in scattering
energies which is too small to be resolved on a nanosecond timescale, 
and the contribution of higher partial waves. The latter becomes
particularly prominent in the presence of resonances.
Fig.~\ref{fig:thermav} compares calculations including both singlet
and triplet channels (a) to those for the triplet component only (b)
in order to highlight the role of shape resonances.
For $^{87}$Rb, shape resonances are observed at
$\sim$160$\,\mu$K ($J=2$) and at $\sim$430$\,\mu$K ($J=6$) in the
singlet, and for $J=2$ at $\sim$290$\,\mu$K in the triplet channel. 
The second peak observed in  Fig.~\ref{fig:thermav}a 
at 1100$\,$ps corresponds to the singlet recovery of the
bleach. The shape resonances lead to a much larger weight of the
singlet contribution in the thermal averages than in the pure
state $s$-wave calculation.
Since probe absorption in the singlet channel occurs at larger
distances than for the triplet channel, cf. Fig.~\ref{fig:scheme}, 
the recovery of the bleach is observed at later times.
This observation opens up the perspective of analysing pair
correlation functions in coupled channels scattering near a resonance
where tuning an external field through the resonance will modify the
respective weight of the channels.
We emphasize that this novel pair correlation spectroscopy is possible
even in the presence of, e.g., three-body losses as long as the
decay occurs on a timescale larger than a few nanoseconds.

\paragraph*{Mapping out the pair correlation function}
Pump-probe spectroscopy, a well established tool in chemical physics,
allows for retrieving the amplitude and phase of a wave function
\cite{OhmoriPRL06}.  In the present context we can retrieve the pair
correlation density operator $ \rho(r,r',t)$ with
$\rho(r,r';t)=\Phi(r;t)\Phi^*(r';t)$ for a BEC.
This is based on Eq.~(\ref{eq:window}), where probe absorption
corresponds to a position measurement with finite resolution:
Different central frequencies, $\omega_P$, define the position that is
measured, and the difference potential, $\Op \Delta(\Op{r})$,
together with the pulse duration, $\tau_p$, control the resolution.
These measurements resolve the amplitude of the pair
correlation, $|\Phi^*(r';t)|^2$. 
The phase information is obtained by chirping the probe pulse which
corresponds to a momentum measurement~\cite{Jiri00}.
The window operator then defines a finite resolution measurement in
phase space~\cite{new}.
Collecting the expectation values for a sufficiently large set of
window operators with different positions/frequencies and
momenta/chirps corresponds to  
quantum state tomography of $\rho(r,r';t)$~\cite{Tomography}.

\paragraph*{Conclusions}
Pump-probe spectroscopy unravels directly many-body pair correlations
in dilute Bose gas.
Existing experimental setups \cite{SalzmannPRL08,McCabePRA09} need to be
only slightly modified to implement our proposal. In particular,
transform-limited pulses of about $1\,$cm$^{-1}$ bandwidth are required
for detection of the probe absorption via molecular ions. 
Spectral features on a scale of less than $1\,$cm$^{-1}$
can be resolved for pump-probe delays of a few nanoseconds.
Pump-probe spectroscopy of the pair correlation dynamics
allows to capture transient states of ultracold gases
such as collapsing condensates. Moreover, it can be
combined with static external field control. For example,
tuning a magnetic field close to a Feshbach
resonance may enhance the pair density at short and intermediate
distances \cite{PellegriniPRL08}. The resulting coupled channels pair
correlation function can be mapped out despite the finite lifetime of
the resonance. 
Future work will consider shaped pulses. Once picosecond pulse
shaping becomes available,  
the full power of coherent control can be employed to study pair
correlation dynamics.

\begin{acknowledgments}
  \paragraph*{Acknowledgements}
  We are grateful to F. Masnou-Seeuws and P. Naidon
  for many fruitful discussions, to our referees for helpful
  comments and to the 
  Deutsche Forschungsgemeinschaft for financial support. 
\end{acknowledgments}


\end{document}